\documentclass[prl,aps,twocolumn,showpacs]{revtex4}
\usepackage{graphicx,color}
\usepackage{dcolumn}
\usepackage{bm}
\usepackage{color}
\usepackage{amssymb}
\usepackage{subfigure}

\newcommand{\beq}{\begin{eqnarray}}
\newcommand{\eeq}{\end{eqnarray}}

\pdfoutput=1

\begin{document}

\title{
Bourdieu Dynamics of Fields from a Modified Axelrod  Model
}
\author{C\'assio Sozinho Amorim}
\affiliation{Department of Applied Physics, Nagoya University,
Nagoya 464-8603, Japan}
%\email{amorim@rover.nuap.nagoya-u.ac.jp}
\date{\today}

\begin{abstract}
Pierre Bourdieu discussed how an individual's taste relates o his or her social environment, and how the classification of \emph{distinct} and \emph{vulgar}, among others, arises from at the same time as shapes this taste  in his work called \emph{La Distinction}. 
Robert Axelrod created a computational model with local convergence and global polarization properties to describe the dissemination of culture by simple selective interactions.
In this letter, Axelrod model is modified, while holding to the same original principles, to describe Bourdieu theory.
This allows to analyze how the dynamics of society's tastes and trends may vary with a simple approach, considering social structures and to understand which social forces are crucial to change dynamics.
Despite the relative simplicity, the present approach clarifies symbolic power relations, a relevant issue for understanding power relation both on large as well as on small and localized scale, with impact on activities ranging from daily life matters to business, politics, and research.
This model sheds light on social issues, showing that a small amount of conflict within a class plays a central role in the culture dynamics, being the major responsible for continuous changes in distinction paradigms. 
\end{abstract}

\pacs{05.10.-a, 89.65.-s, 87.23.Ge
}

%05.10.-a	Computational methods in statistical physics and nonlinear dynamics
%89.65.-s	Social and economic systems
%75.10.-b	General theory and models of magnetic ordering
%89.75.Fb Structures and organization in complex systems
%87.23.Ge Dynamics of social systems
%05.50.+q Lattice theory and statistics (Ising, Potts, etc.) 

\maketitle

%\section{Introduction}

{\it Introduction --- }
It is often said that ``there is no accounting for taste.'' Nevertheless, an individual taste and behavior is often classified as either good or bad in many ways: %on a plethora of instances:
 sophisticated or vulgar, developed or crude, exquisite or dull, elegant or rough -- to name a few. %This taste is also commonly used to classify individuals.

Pierre Bourdieu analyzed in deep how French society displayed different tastes according to social classes, and how a person's taste is also used to classify this person\cite{Bourdieu}. %ref
This social classification permits to identify and classify individuals according to dominated and dominant class, and moreover the dominant fraction of the dominant class and the dominated fraction of it.
This dominant class takes the so called \emph{distinct} cultural traits  chosen by rarity through cultural monopoly, obtained by means of economic power, and tag the dominated-class traits as \emph{vulgar}. Overdone pretension of middle class to display superior distinct aspects is not seen as ``natural distinction,'' also being regarded as vulgarity.% This interaction and exchange between cultural traits happens within the dynamics of the (social) fields in question.

Robert Axerold, on his side, proposed a mathematical model for computers to describe the spread of culture, in principle unrelated to Bourdieu's theory\cite{Axelrod}. %ref
Axelrod's model works on a lattice with $n$ features per lattice point, each feature having $k$ possible traits to assume. Interaction happens between nearest neighbors with probability proportional to the number of equal traits between the two sites. When an interaction takes place, one of the differing traits of one site is set equal to the other site's trait. Tables \ref{axel1} and \ref{axel2} illustrate the situation. 
 Although many models exist for a diversity of purposes \cite{TT06, Galam, CFL09, LNL92, SS00, Galam05, CFGG06}, %refs here!
Axelrod model has a good balance between simplicity and realism, and has been extended on several ways to include media effect and other changes\cite{PF11, GJ05, RM10, SYI01, AECKHM06, MCD07, ACM12, DPM09}, with many researches about its characteristcs available \cite{CMV00, KETM03, GCFB11, ACM14, PKK13, Fontanari10, SRKK14}. %ref
 An interesting change is the introduction of a repelling effect in the model \cite{DPM09}. %ref 
 Since the Axelrod interactions  are limited to copying a neighbor's characteristic and therefore aligning more and more to it, one can say that only ferromagnetic-like orders arise.
Such kind of interaction can be justified by affirming that people interact more frequently with those who are culturally similar to them, tending to agree culturally. 
%While interesting and useful as is, an antiferromagtetic-like interaction could also be seen as possible if taken a broader picture: a person tends to dislike certain cultural aspects and avoid it.
But when looking at cultural interactions from an angle that encompasses both like-and-hate aspects,  we come to Bourdieu's taste theory.

In this letter, we take an approach based on a layered Axelrod model, where the layers interact mainly according to an attracting (repulsive) Axelrod model within it (between them), representing social affinity (avoidance). A dynamics that can be related to Bourdieu's field dynamics is observed to arise and the results show how small  perturbations lead to essentially different behaviors, indicating how slight in-class conflicts may fundamentally change the results. This can also be regarded as a more realistic cultural drift effect, adopting agent interactions as a source of drift, instead of a random step. % are discussed under both application and theoretical points of view, with special account to the role played by the AF part in social phenomena.

\begin{table}[hbt]
\begin{tabular}{rcccl}
%\hline
& \ldots& 08441 &\ldots&\\
\ldots & 29364 & \underline{29462} & 97083 & \ldots\\
& \ldots& 33942&\ldots&
\end{tabular}
\caption{An example of a lattice for Axelrod model. The underlined numbers correspond to a lattice point with $n=5$ features, each one assuming traits  valued from 0 to 9. The nearest neighbors are also shown, on a two-dimensional square lattice. Any of the other four points in this table can be chosen to interact with the underlined one.}
\label{axel1}
\vspace{10 pt}
\begin{tabular}{rcccl}
%\hline
& \ldots& 08441 &\ldots&\\
\ldots & 29\underline{3}64 & 29\underline{3}62 & 97083 & \ldots\\
& \ldots& 33942&\ldots&
\end{tabular}
\caption{If we choose in table \ref{axel1} the point to the left to interact, there is a 60\% chance of interaction which, in case it happens, could lead to change of the central trait from 4 to three, underlined here, to adjust to the neighbor's value.}
\label{axel2}
\end{table}

{\it Axelrod Model and extension---}
The traditional, attracting  Axelrod model can be best described by a step-by-step algorithm:

\begin{enumerate}
\item Choose randomly a site $i$ and a neighbor $j$.
\item If the sites have $n'$ features with the same trait value among $n$ total features, they interact with probability $n'/n$.
\item If an interaction takes place, a random feature $\alpha$  from the differing $n-n'$ features of the site $i$ assumes the same trait of $\alpha$ feature of the site $j$.
\end{enumerate}

This describes one step of the model, which is iterated many times over, and is exemplified in tables \ref{axel1} and \ref{axel2}. The interaction probability $p_{i,j}$ in step 2 can be set as 

\begin{eqnarray}
p_{i,j}=\frac{1}{n}\sum_{t=1}^n v_{i,t}\cdot v_{j,t} \equiv(1-d^A_{i,j}) \, ,\\
v_{i,t}\cdot v_{j,t} \equiv \delta_{v_{i,t}, v_{j,t}}\, ,
\end{eqnarray}

with ${v}_{i,t}$ the $t$-th component of the vector $\boldsymbol{v}_i$ holding as components the features of the $i$-th site, and we call $nd^A_{i,j}\equiv \sum_t (1-{v}_{i,t}\cdot {v}_{j,t})$ the Axelrod distance between two sites, where $d^A_{i,j}=1$ ($d^A_{i,j}=0$) stands for completely orthogonal (aligned) sites. We shall also define the distance between two sites $i,j$

\begin{eqnarray}
d^2_{i,j}=A^2\sum_{t=1}^n(v_{i,t} - v_{j,t})^2\, ,
\label{dist}
\end{eqnarray}

where $A$ is a normalizing constant, for later use.

%Under a Hamiltonian representation, one can write it as
%
%\begin{eqnarray}
%H=-\sum_{<i,j>}\sum_{t=1}^{n}\boldsymbol{v}_{i,t}\cdot\boldsymbol{v}_{j,t},
%\end{eqnarray}
This algorithm will be used  with two changes for an attracting (A) interaction. First, we take periodic boundary conditions, abandoning finite-size effects presented in ref. \cite{TT06} and adopting a toric topology for our two-dimensional square lattices. Second, in the last step, instead of a straightforward copy of the trait value, we assume that this happens with a certain probability $p(k,k')$, where $k$ ($k'$) is the former (new) trait value. In principle, it is also assumed that higher values are more \emph{expensive} to be adopted, hence less probable. The simplest probability function is of the form $e^{-\beta k'}$, ignoring the initial value, but this probability does not show that an individual displays large resistance to big changes, what should make transitions between close values be relatively easier than those between values far apart, even if from a very expensive (improbable) to a very cheap (probable) state, as long as the former state can be afforded. This relation leads to a probability 

\begin{eqnarray}
p(k,k') \propto e^{-\beta |k'-k|},
\label{prob}
\end{eqnarray}

that still keeps transitions between traits with close cost quite probable, and creates a resistance for exchanging far away values. Actually, it is worth mentioning that the probability taken here is better described as a functional of a probability density function (PDF), which is discussed in the supplemental material.%shall be discussed in more details later in this paper.

For a repelling (R) Axelrod model, again step 3 in the above algorithm is changed, where instead of assuming a neighbor's trait, one shared trait is changed to a random different value with probability $p(k,k')$ given by Eqn. (\ref{prob}). From here on, it will be shown that this A/R altered Axelrod algorithm is enough to reproduce and explain some  aspects of Bourdieu distinction on a quite clear basis, under the right assumptions.

\begin{figure}[tbh]
\includegraphics[width=80 mm]{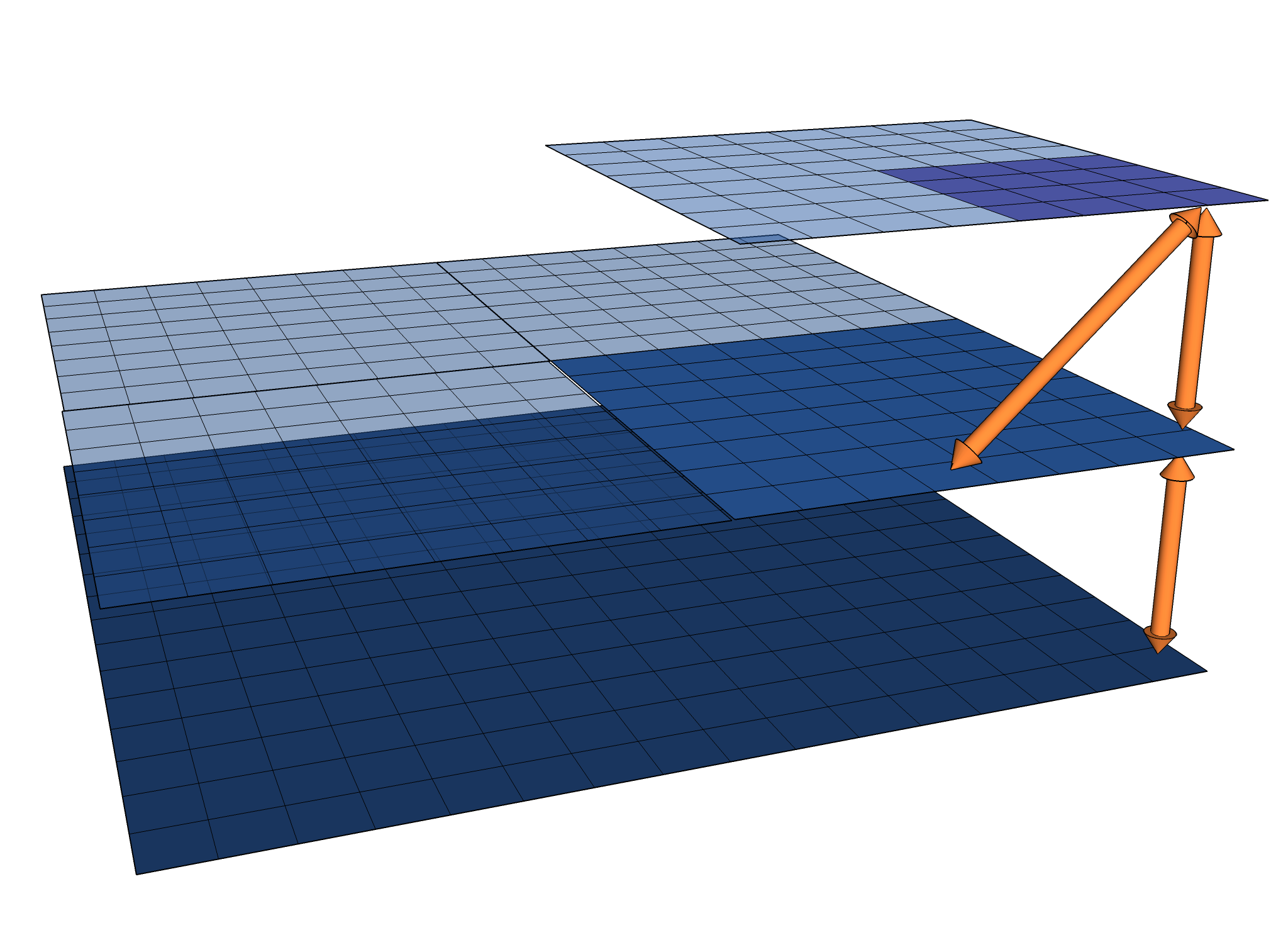}
\caption{Tri-layer lattice model. The bottom layer ($l=3$) represents the dominated class, while the top layer ($l=1$) stands for the dominant fraction of the dominant class, and the middle ($l=2$), the dominated fraction of it. Periodic boundary conditions are taken differently for each layer, with each period shown in solid colors, and its replication transparent. Arrows represent possible interlayer interactions, taking into account layers period.}
\label{model}
\end{figure}

Consider a tri-layer stacked system as shown in Fig. \ref{model}, where each layer represents a different sector of social classes. 
The bottom-most layer represents the dominated class, while the top-most should stand for a dominant class  and the intermediate layer would be the dominated (pretentious) fraction of the dominant class. 
According to Bourdieu, the \emph{habitus}, i.e. the set of environment, customs, lifestyle, dispositions, etc. of a social group, molds an individual's tastes and binds it to a certain class, while generating a resistance toward other classes' tastes. We therefore set intra(inter)-layer interactions between nearest neighbors to be A(R) Axelrod based, reproducing such affinity (avoidance), after some iterations restricted to intra-layer interactions for the {\it habitus} setup. 
Furthermore, inter-layer interactions are supposed to be much less present as a sign of weakness, happening less frequently. Here, one fifth of the present interactions are taken to be inter-layer type, four times less than intra-layer ones.

Inter-layer interactions occur in two ways. The first, as all the intra-layer interactions, acts between immediate neighbors, i.e. lattice points immediately above or below the layer. For the second, between non-neighbor layers, interactions occur with the (rounded) mean  features of the other layer, a sort of mean-field approach. The meaning of such interaction is that, despite ignoring the existence of any direct interaction between the top-most and bottom-most layers (dominant and dominated classes), the richest stratum has means to influence the whole society by dictating paradigms and stereotypes, holding cultural monopoly over some expressions, while depreciating cultural manifestations of the dominated instance. 
It is worth stressing that this could, in principle, by manipulation of media, influence other layers under an A Axelrod interaction, but this effect will not be addressed directly here for clarity.
Therefore, such mutual exclusion can only be perceived as a mean character of the farthest layer, which can be given by a vector $g_l$ with each trait value as a component, where $l$ is the layer index ($l=1,2,3$) ordered top-down. For this tri-layer model, as the middle layer has equal probability to interact with each other layer, so the extremity layers have equal probability to interact with the middle one, or with the mean trait vector $g_l$ of the opposite extremity.

Up to this point, no specific difference between the strata which compose the system has been stated, leaving full symmetry between dominant and dominated classes. This symmetry is broken in two ways. The first consists of a change in the probability function for each case. This is simply achieved by setting $\beta$ in eqn. (\ref{prob}) to $\beta_l=lc$, where $l$ is the layer index and $c$ a constant. For the calculations in this work, $c=0.25$. This makes layer 1 hold access to about twice more trait values than layer 3 initially, holding control on some cultural features.
 The second difference is inserted in the period of the boundary conditions. Layer 2 (3) is supposed to have a $s$ times larger period than layer 1 (2), shown in solid colors in Fig. \ref{model}.
 %%%%%%%%%%%%%%%%%%%%%%%%%%%%%%%%%%%%%%%%%% what to say?
  %This geometric progression of the period is again a simple but meaningful approach, as differences in society arise commonly in this way, e.g. 10\% of population on a dominant position with 1\% on a fully dominant instance. 
  This change in period implies more global similarity among dominators, with more divergence amidst dominated agents, but is justified rather by a smaller population with similar \emph{habitus}. Therefore, this model can be seen as a pyramid of three layers, with each one having the size of its period. Another important  meaning is also embedded in this assumption. Longer periods as taken in this model are also a way to effectively adjust social strain on each layer. Making an analogy with waves, for a general wave equation, (phase) velocity varies squarely with the stiffness. 
  %On a string $v^2=\frac{T}{\mu}$, where $T$ is the tension and $\mu$ a (constant) linear density; 
  %On a cotinuum medium, waves assume different velocities, with $v^2 =\frac{B}{\rho}$, where $B$ is the bulk modulus,  and $\rho$ the density. 
  In general, the higher the tension or stress in a system, or the higher the stiffness, one can then expect longer wavelengths due to higher velocity for same frequencies. This principle is taken to attribute a longer period also for classes under higher social strain.

\begin{figure}[bht]
\includegraphics[width=80mm]{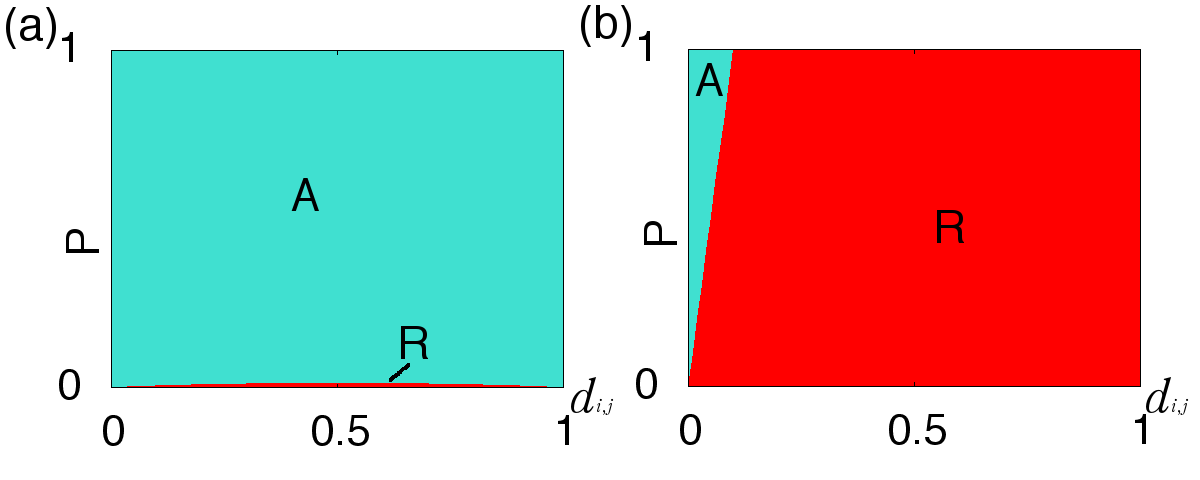}
\caption{Probability of different types of interactions for (a) intra-layer and (b) inter-layer cases. The red (green) region indicates R (A) interaction.}
\label{d-p}
\end{figure}

Now, as there is no special reason to take all the interactions to be fixed as A/R, we add the possibility of both of them as a function of distance $d_{i,j}$ in Eqn. (\ref{dist}). We take this distribution to be as in Fig. \ref{d-p} and verify how this changes the system evolution on average.

{\it Simulation Results ---}
%Results from a single run are shown in <here>.%add it 
Here, each lattice point is supposed to represent an agent with $n=6$ features, each feature assuming a trait  $k\in \{0,1,\ldots, 9\}$. For the formation of the \emph{habitus}, i.e. the background intrinsic characteristics associated with a class, random traits are attributed to each point, according to the probability distribution of the layer it belongs to, and only intra-layer interactions are turned on for a million iterations, corresponding to 1\% of the simulation.. Therefore, each lattice point starts at a random point in a $\mathbb{Z}_{10}^{\otimes 6}$ space, and each layer tends to end up in one or close to one point in it (several, in case of domain formation or poor convergence). As the initial state in each layer is fully random, the initial mean taste $\boldsymbol{g}_{l}^{\circ}\equiv \langle \boldsymbol{v}_{i}\rangle$, defined as a vector with components equal to each mean feature of a layer $l$, is used to define an order parameter

\begin{eqnarray}
D_l = \sum_{t} (g_{l,t} - g^{\circ}_{l,t})^2,
\label{layer_dist}
\end{eqnarray}

with $g_{l,t}^{(\circ)}$ being the $t$-th component of the vector $\boldsymbol{g}_l^{(\circ)}$. Yet, this order parameter fails to be an index on how high (low) are the values of each layer, for which an intensity parameter is introduced,

\begin{eqnarray}
G_l = \sum_{t} (g_{l,t})^2.
\label{layer_norm}
\end{eqnarray}

By looking at both $D_l$ and $G_l$ one can tell (a) how well ordered is a class under inter-layer repulsion and (b) how ``exuberant'' is such class. After the {\it habitus} is formed, inter-layer interactions are turned on. Two different cases are discussed. The first, with pure A (R) intra(inter)-layer interactions, and the second where the choice of interactions type is made according to the probability shown in Fig. \ref{d-p}. Fig. \ref{d-p} (a) R region obeys a quadratic distribution with maximum probability of 0.5\% at $d_{i,j}=0.5$, while Fig. \ref{d-p} (b) has a linear border starting at the origin and finishing at $d_{i,j}=0.1$. %The average evolution of such evolution is seen in Fig. \ref{amop}.
%As a general tendency, the inter-layer interaction compels the whole system to take higher values of traits. In other words, the more common lower values are vulgarized and avoided by the whole population. Nevertheless, local observation of layers shows an approximation of layer 2 to layer 1, while layer 3 changes more locally, with more diversity. Layer 1 has a sort of trend making power, agreeing within it and changing the stablished standards from times to times, also due to influence of layer 2. Here, the probability function adopted should ideally be adjusted according to plateaus layer 1 order parameter, since this dynamics play important role in social organization, but this effect is being ignored for simplicity. 

\begin{figure}[hbt]
\centering
\subfigure[$G_l$]{
\includegraphics[width=40 mm]{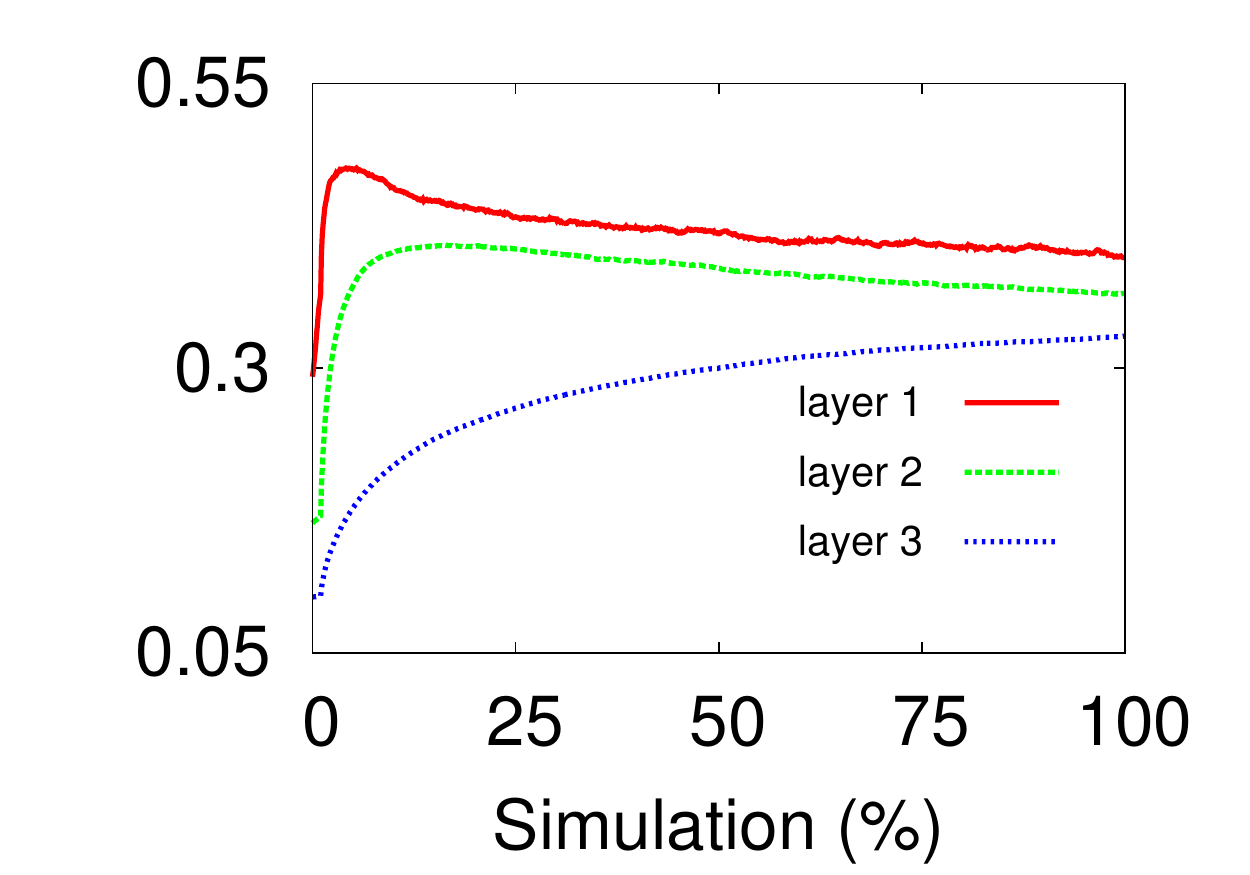}}
\subfigure[$D_l$]{
\includegraphics[width=40 mm]{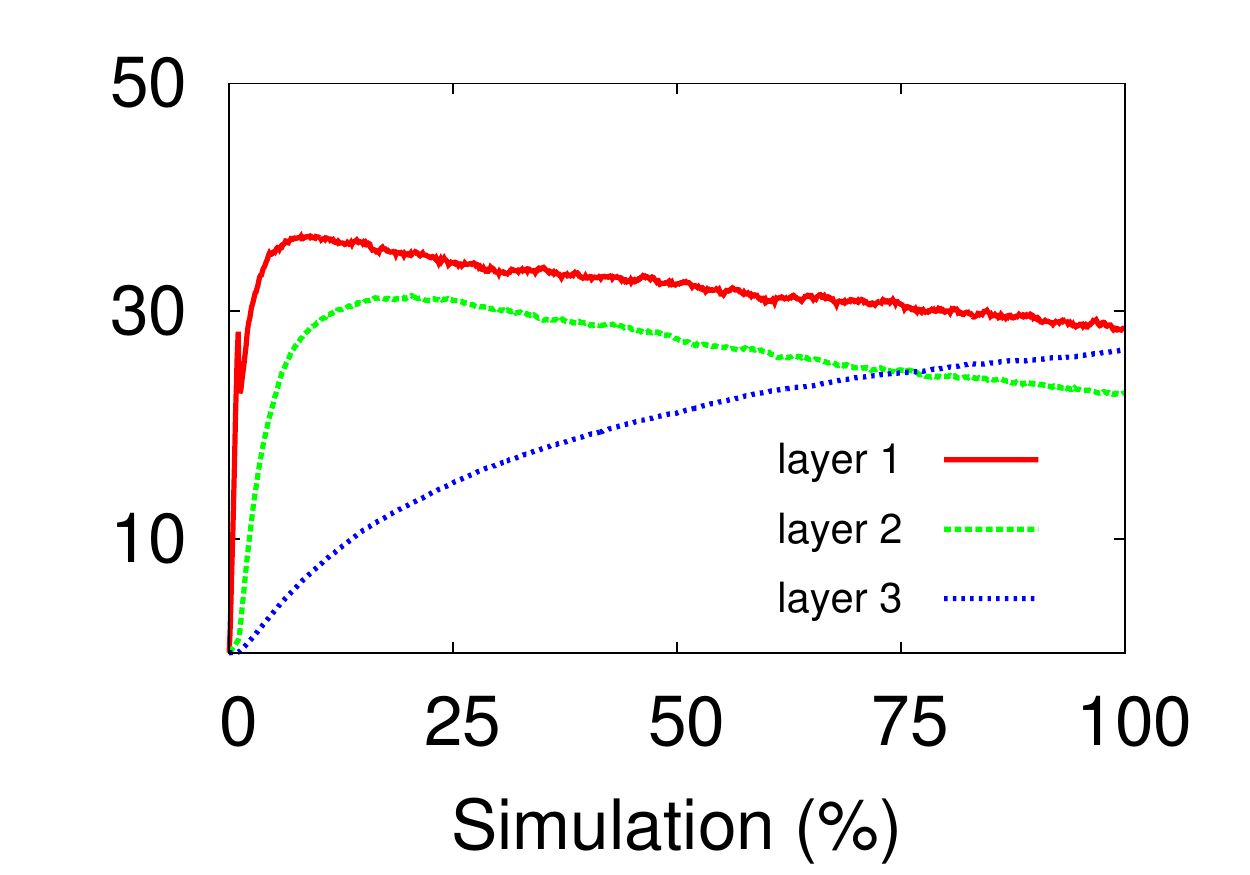}}
%\caption{$G_l$ and $D_l$}
%\label{amop}
%\end{figure}
%\begin{figure}[hbt]
%\centering
\subfigure[$G_l$ w/o perturbation]{
\includegraphics[width=40 mm]{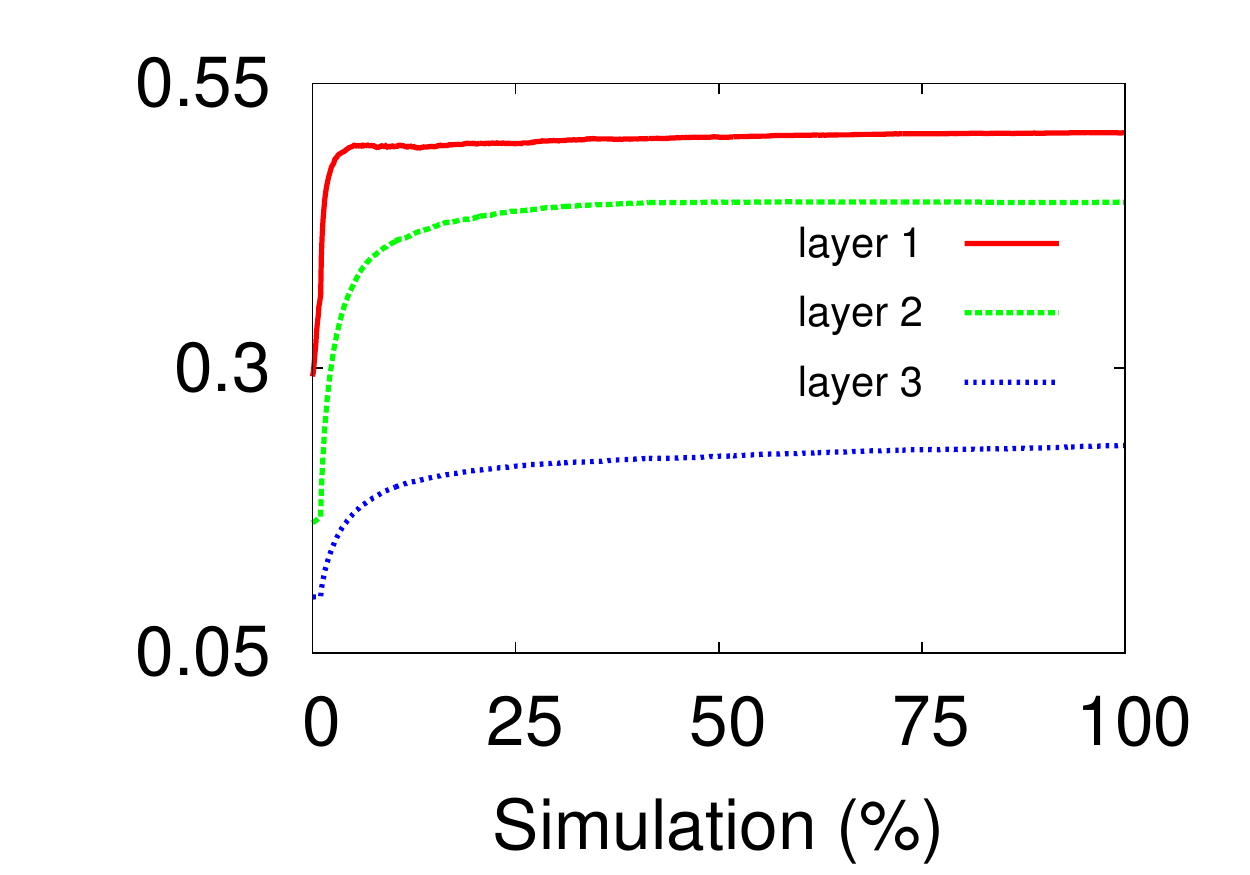}}
\subfigure[$D_l$ w/o perturbation]{
\includegraphics[width=40 mm]{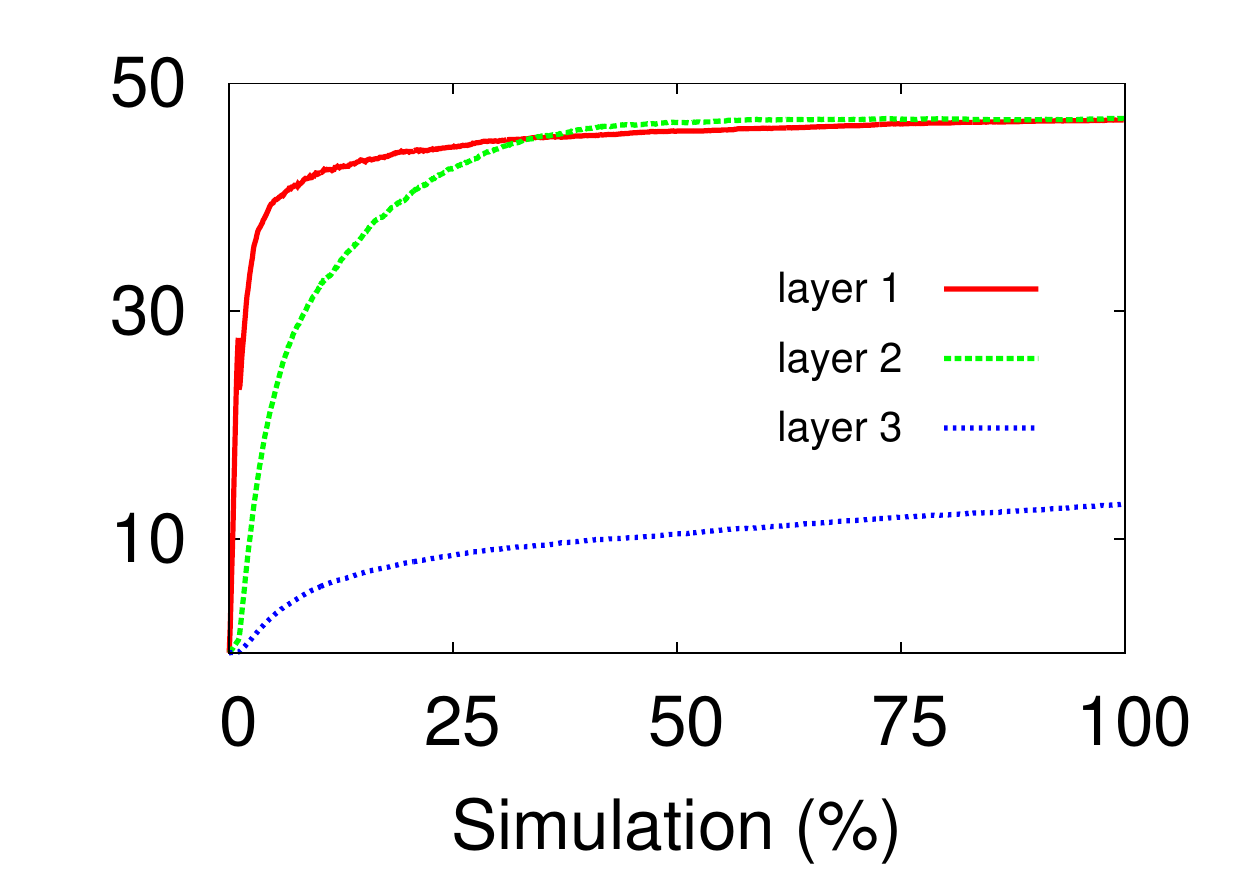}}
\caption{$G_l$ and $D_l$ average evolution within a simulation obeying the interaction distribution in Fig. \ref{d-p} in (a) and (b) and with pure A (R) in-(inter-)layer interactions in (c) and (d).}
\label{amop}
\end{figure}

%We can now come back to Bourdieu's theory and verify how does it relate with the present results.  Notably, it is interesting to verify how the habitus is molded and changed under the present social assumptions, i.e. how a layer (or social class) varies, and what the average final $G_l$ and $D_l$ tells us.

\begin{figure}[hbt]
\centering
\subfigure[Layer 1]{
\includegraphics[width=27 mm]{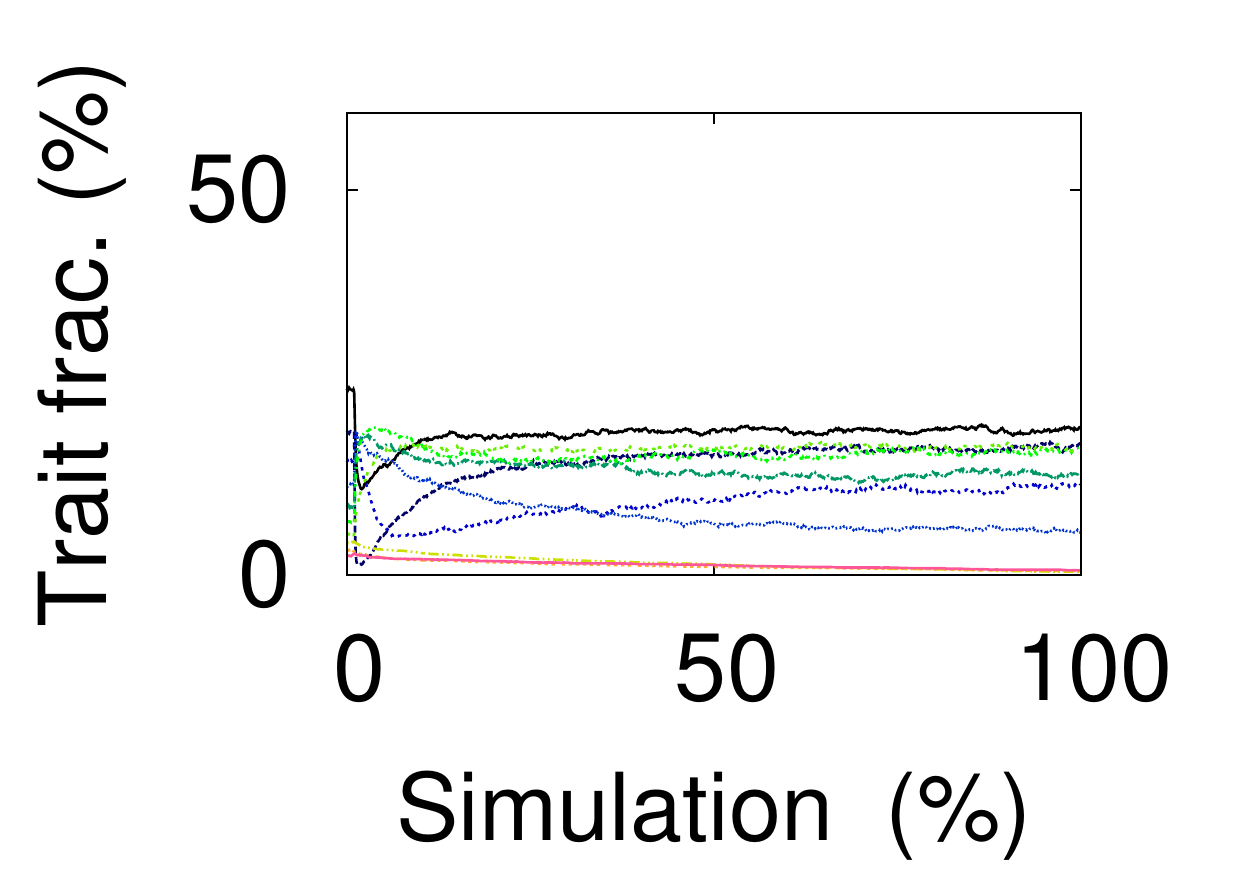}}
\subfigure[Layer 2]{
\includegraphics[width=27 mm]{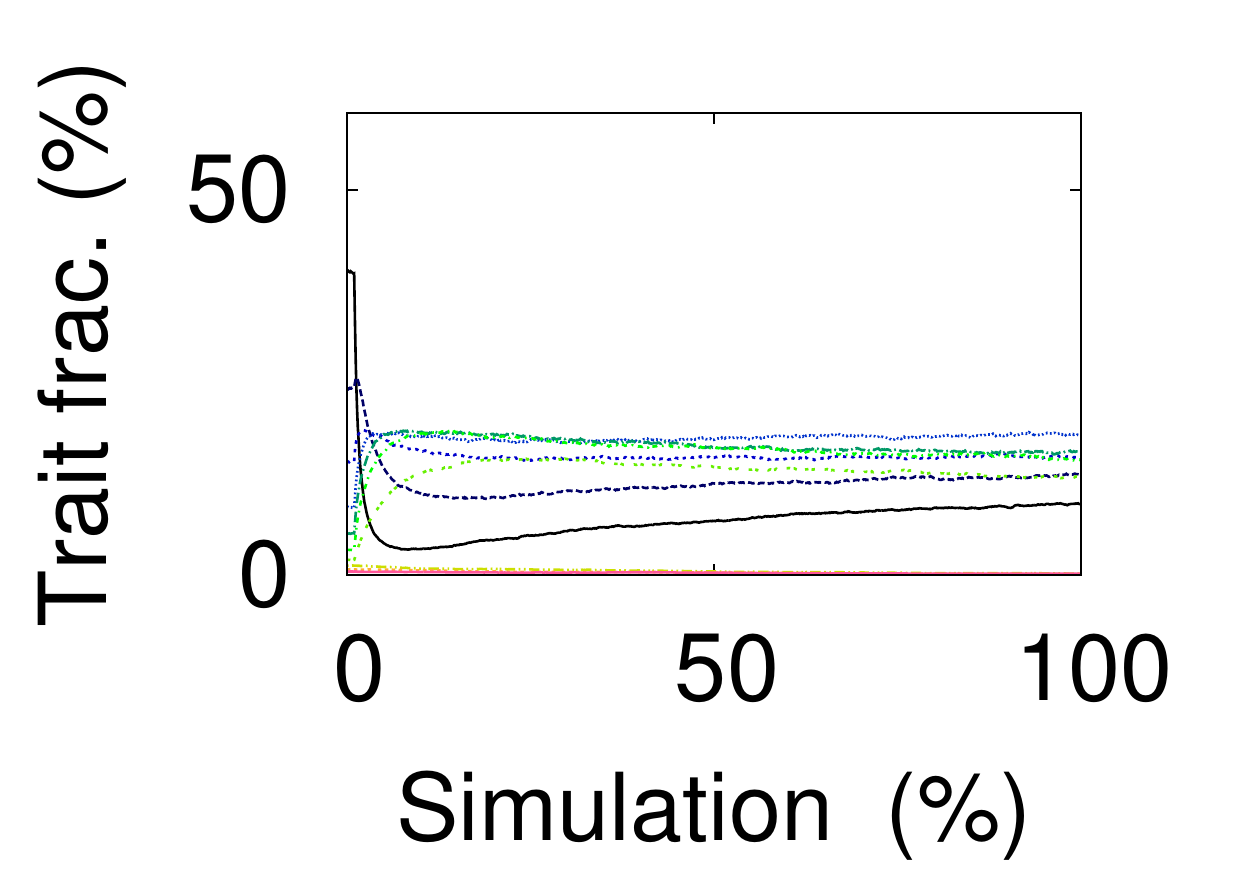}}
\subfigure[Layer 3]{
\includegraphics[width=27 mm]{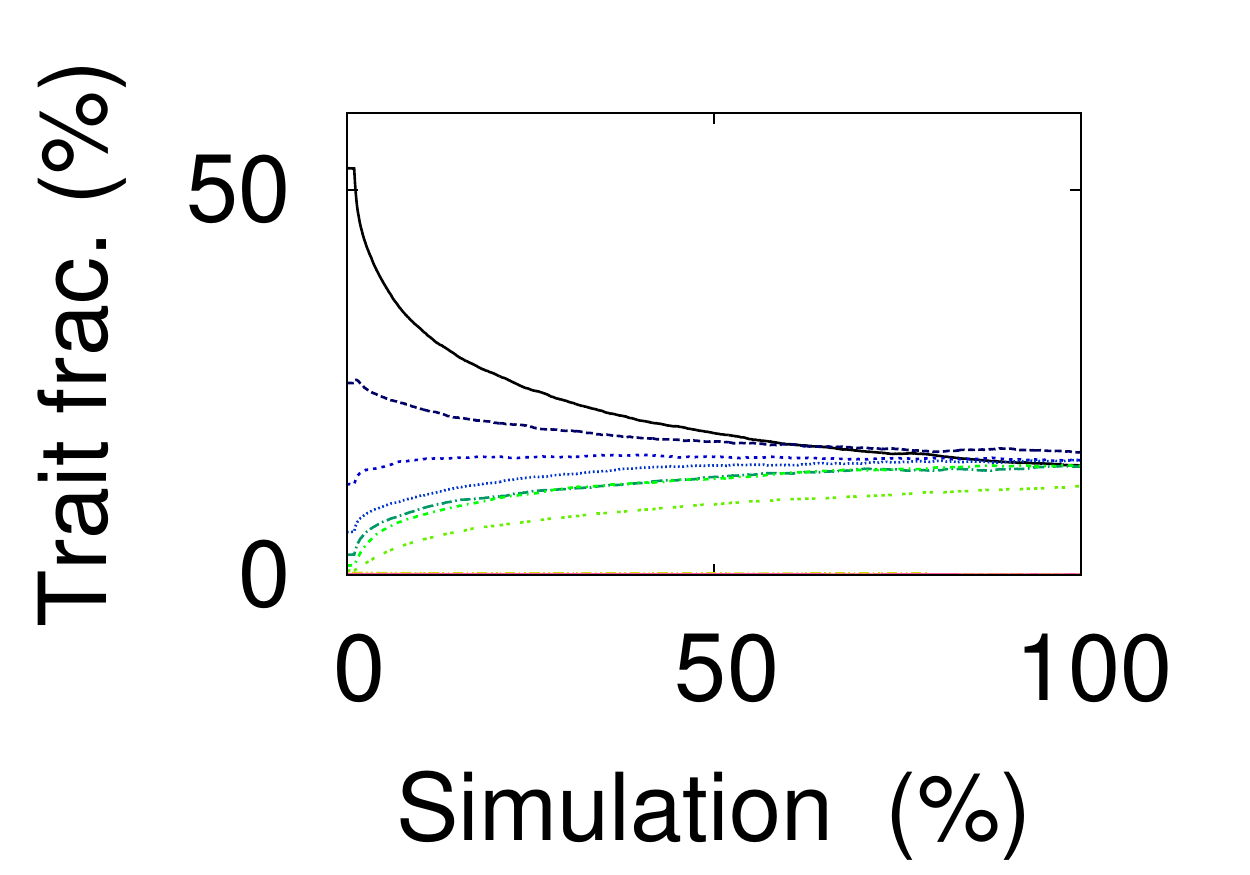}}
\caption{Average trait evolution for layers (a)1, (b) 2, (c) 3. Brighter colors indicate higher traits.}
\label{ch}
\end{figure}

Fig. \ref{amop} gives us $G_l$ and $D_l$ evolution averaged over 1000 trials. Panels (a) and (b) of Fig. \ref{amop} shows the case following the disturbances introduced in Fig. \ref{d-p}, whereas panels (c) and (d) use only A (R) interactions within a (between) layer(s).  $G_l$ has an abrupt general raise, as a sign of avoidance of similarities, which initially happens mostly among highly probable low values. Hence, this can still be seen as part of the setup of the system, and we must look at how does it evolve from there on. The lowest layer displays a continuously increasing $G_l$ and $D_l$. The later can be understood as an increasing order of the dominated class, which does not possess random culture features, but rather limited traits as limited by the dominant classes. On the numerical side, it is also important to remember that the definition of $D_l$ sets a relatively low standard ($g_l^{\circ}$) for the lowest stratum, which makes it easier to be pushed to higher values. The same  is valid for $G_l$, with the difference that how much it can grow depends strongly on how intra-layer interactions are defined. For the two dominant classes, a rather accentuated drop of $G_l$ is observed in Fig. \ref{amop} (a), not present in (c). This can be understood under Bourdieu's argument of them looking for distinction, with the top-most layer legitimating a ``natural distinction'' against a ``pretentious distinction'' often exaggerated, which can in turn be seen as vulgar. 
%But the stronger the intra-layer repulsion, the closer $G_l$ tends to approach each other, also with crossings due to oscillation, and convergence of values among the dominated layers. 
%Therefore, no convergence of these layers is seen, but a repulsion or crossing instead. 
Similarly, $D_l$ also drops, with the dominant classes exploring the available rarity to them less orderly, as a result of their intrinsic pursue for distinction. These results can also be confirmed from Fig. \ref{ch}, which shows the average traits evolution per layer. The lowest layer keeps lower (darker) traits, lesser present in the middle layer. These raise on the top layer, which again can be seen as an avoidance of exaggerated pretension, or a pursuit of the ``natural distinction,'' while still keeping higher (brighter) traits. Notice that this does not manifest which features assume such traits, what is still generally orthogonal between layers.

Now, it is interesting to observe that this effect depends strongly on how the probability of intra-layer R interactions, i.e. in-class conflict, is set. While a full A order gives us the simple result shown in Fig. \ref{amop} (c) and (d), the stronger the conflict within the layers, the closer they tend to become to each other. It is interesting to see that the A interaction portion between layers does not display the same behavior, leaving almost no difference even if not present at all. This leads us to say that small conflicts within a certain class holds more ``self-destructive'' power, in the sense that the class reshapes to gain more similarity to others, than different classes trying to dictate or influence an appropriate behavior. That is, an agent confronted with the dilemma of aligning its cultural capital to its surrounding field or with a different field will eventually be caught by the highest frequency interaction, i.e the alignment with its own group. On the other hand, a small disturbance in this field will introduce several new cultural trends in it which, eventually, one of them will spread around reasonably fast, until a new concept arises from a new conflict, scattering the propagating tendency. The more frequently this happens, the less uniform a layer becomes, leading to an ill-defined distinction. Therefore, this suggests us that Bourdieu's distinction defined within the dynamics of fields is sustained by a very small, perturbative conflicts within these fields, but that are kept dilute to maintain  dominance and symbolic power. Disputes within a class show a bigger role in this class disrupt than other external factor, as far as they are considered here. 

{\it Conclusion ---}
We conclude that Bourdieu's dynamics of fields for legitimating symbolic power and cultural distinction can be seen as a social phenomenon which can be described with Axelrod culture dissemination model on a stacked toric lattice with  competitive attractive-repelling interaction. Moreover, without a little influence of conflicting in-layer interaction, the resulting frozen state becomes unnatural and less dynamic under Bourieu's point of view, which allows us to understand the small portions of conflict within a social class as a crucial factor for the evolution of the underlying classifying and classified cultural traits. 

\vspace{5ex}

\noindent
The author is grateful to D. Nardi and B. Hayashi. %Whom?God?
for fruitful discussions. This work was supported by Japanese Government (MONBUKAGAKUSHO:MEXT) Scholarship for 2013 ({\it gaikokujinryuugakusei}).  %What?Holy Spirit?

\bibliography{socio}

\appendix

\setcounter{equation}{0}
\renewcommand{\theequation}{S\arabic{equation}}

\section{Supplement}
It is worth to comment on the probability function presented in the main text. In more general terms, considering also the possibility of continuous trait values, the probability governing the transition from one value to another can be given by a functional of the form

\begin{eqnarray}
\mathcal{P}[p] = \int \mathrm{d}\kappa\, p(k,\kappa,t)K(k',\kappa,t),
\label{prob_functional}
\end{eqnarray}

with $K$ being the kernel of the functional, $p$ a PDF, and $t$ indicating time. In the present calculations, $K(k',\kappa,t) = \delta(\kappa-k')$ and $p(k,\kappa,t) = e^{-\beta |k-\kappa|}$ as the simplest case of one value changing to another well-defined value and only one, ignoring transitions to other values close to the surroundings of the target value. Actually, since only discrete values are taken here, the integration can be substituted for a summation, and the delta function for a Kronecker's delta. Time dependency is also disregarded, but it is worth noticing that this implies limitations. Bourdieu presents his work on dynamics of fields  considering a very large time scale, comparing society in completely different periods of times, centuries away from each other. Under this circumstance, the definitions of \emph{distinct} and \emph{vulgar} change, even allowing  certain inversions of concepts. Since rarity is an important index for the elite taste, change of rarity in time should be included in eqn. (\ref{prob_functional}) to cover the whole dynamics. Nevertheless, such complications would only be an obstacle for a first evaluation, obscuring the central aspects for interpretation. Hence only the static, simple limit of a constant probability distribution $p(k,k')$ is taken. Ideally, we may quote Bourdieu in ref. \cite{Bourdieu} saying ``The demand which is generated by this dialectic is by definition inexhaustible since the dominated needs  which constitute it must endlessly redefine themselves in terms of a distinction which always defines itself negatively in relation to them.'' Therefore, an oscillatory behavior is naturally expected to happen, but we suppress this effect in the present considerations. On the other hand, this still claims for the initial drop observed in the indices in the main text, which must become the trigger for the expected oscillations once Bourdieu's feedback is taken into account.

\end{document}